# New measurement of neutron electric dipole moment

A.P. Serebrov[1)], E.A. Kolomenskiy, A.N. Pirozhkov, I.A. Krasnoshekova, A.V. Vasiliev, A.O. Polyushkin, M.C. Lasakov, A.K. Fomin, I.V. Shoka, V.A. Solovey, O.M. Zherebtsov

*B.P. Konstantinov Petersburg Nuclear Physics Institute of National Research Centre "Kurchatov Institute", 188300 Gatchina, Leningrad region, Russia*

P. Geltenbort, S.N. Ivanov, O. Zimmer

*Institut Max von Laue – Paul Langevin, BP 156, 38042 Grenoble Cedex 9, France*

E.B. Alexandrov, S.P. Dmitriev, N.A. Dovator

*Ioffe Physical Technical Institute RAS, 194021 St. Petersburg, Russia*

We report a new measurement of the neutron electric dipole moment with the PNPI EDM spectrometer using the ultracold neutron source PF2 at the research reactor of the ILL. Its first results can be interpreted as a limit on the neutron EDM of $|d_\mathrm{n}| < 5.5 \times 10^{-26}$ e·cm (90% confidence level).

Elementary particles may possess a static electric dipole moment (EDM) only if parity (P) and time-reversal (T) symmetries are simultaneously violated and hence via the CPT theorem also the combined symmetries of charge conjugation C and parity. Since CP violation implemented into the Standard Model via the CKM mechanism is known to be able to produce only a much smaller baryon asymmetry than actually observed in the Universe, additional sources of CP violation are needed. These appear for instance in supersymmetric models with multiple Higgs-particles or left-right symmetric theories. Particle electric dipole moments (EDMs) are a sensitive probe of such new physics, complementary to searches for new particles with high energy accelerators. While a recent measurement of the electron EDM [1] seems to challenge the minimal super-symmetric model of electroweak baryogenesis (see Fig. 2 in [2]), further improvements of sensitivity in searches for particle EDMs and notably that of the neutron will be necessary to finally discover the necessary additional CP violation [3].

Most accurate constraints on the neutron EDM value have been obtained using Ramsey's magnetic resonance method applied to polarized ultracold neutrons (UCN) trapped in chambers situated within a static and homogeneous magnetic field [4, 5]. A typical measurement cycle involves filling of traps for 50 s, storage for 70 to 100 s, and emptying for 50 s. After filling, a π/2 magnetic radio-frequency pulse is applied at the resonance of free neutron precession in the static field. A second π/2 rf pulse coherent with the first one is applied after a time interval $T$. Afterwards the trap is emptied with polarization dependent detection of neutrons. Two sequential cycles with different polarities of an electric field applied parallel (antiparallel) to the static magnetic field constitute a basic measurement of the neutron EDM. Its value is obtained from

$$d_\text{n} = \frac{h(N^+ - N^-)}{2(E^+ + E^-)\frac{\partial N}{\partial f}}, \tag{1}$$

where $h$ is Planck constant, $N^+$ and $N^-$ are neutron counts for parallel and antiparallel directions of electric and magnetic fields, $E^+$ and $E^-$ are the corresponding electric field strengths, and $\frac{\partial N}{\partial f}$ is the slope of the resonance curve at the working point defined further below. The statistical uncertainty of the EDM measurement is given by

$$\delta d_\text{n} = \frac{h\sqrt{N^+ + N^-}}{2(E^+ + E^-)\cdot\frac{\partial N}{\partial f}} = \frac{h}{2(E^+ + E^-)\cdot \alpha \cdot \pi \cdot T \cdot \sqrt{N^+ + N^-}} \tag{2}$$

where $(N^+ + N^-)$ is the total number of counted neutrons at resonance frequency, and $\alpha = (N_\text{max} - N_\text{min})/(N_\text{max} + N_\text{min})$ is the visibility of the resonance curve, $N_\text{max}$ and $N_\text{min}$ are counts at the maximum and minimum of the resonance curve.

---

[1] E-mail: serebrov@pnpi.spb.ru

The work reported here employs the differential magnetic resonance spectrometer developed at PNPI [4] (see Fig. 1). One of its peculiar features is a stack of two UCN storage chambers placed in a common magnetic field in vertical direction. Electric fields in the chambers are equal in magnitude but oppositely directed. Chambers and coils for the static and rf fields are located within a four-layer magnetic screen made of permalloy with a dynamical shielding factor of about $10^3$. Control and stabilization of magnetic resonance conditions is maintained with the help of eight quantum self-oscillating magnetometers using optical pumping of cesium atoms. The magnetometer oscillation frequency corresponds to the Larmor precession of the Cs atoms in the static magnetic field and is used to control and stabilize the magnetic resonance for neutrons. Their sensitivity to magnetic field variations is ~ 1 fT for a measurement time of 100 s, as determined by the level of their intrinsic noises. These are negligible compared to the level of magnetic field perturbations inside the shield. A detailed study of the magnetometers in the magnetic screen is presented in [6]. They are placed around UCN storage chambers in pairs symmetric to the plane defined by the central electrode. The whole assembly provides monitoring of the mean magnetic field within the resonance range. As in previous measurements [4] we used the method of stabilization of resonance conditions [7], with electronics upgraded for modern digital signal processing [8, 9]. Division of the average frequency of all magnetometers by the ratio of gyro-magnetic ratios for cesium and neutrons provides the neutron resonance frequency, while by changing the division factor one can sweep through the resonance curve for adjustment of resonance conditions.

Electric fields in the chambers are generated by a high voltage source providing up to 200 kV with polarity reversal of the output voltage without disconnecting the load. The high voltage system was renewed maintaining its construction principle presented in [4, 10] but adopting a more compact and reliable design. The electrically insulating, cylindrical wall of each UCN storage chamber consists of a ring made either of fused-quartz or glass ceramic. The inner surfaces of the rings are coated either with beryllium oxide or Ni/Mo oxide which are good insulators and have a sufficiently high limiting velocity for UCN reflection. The inner surfaces of the electrodes were covered with beryllium.

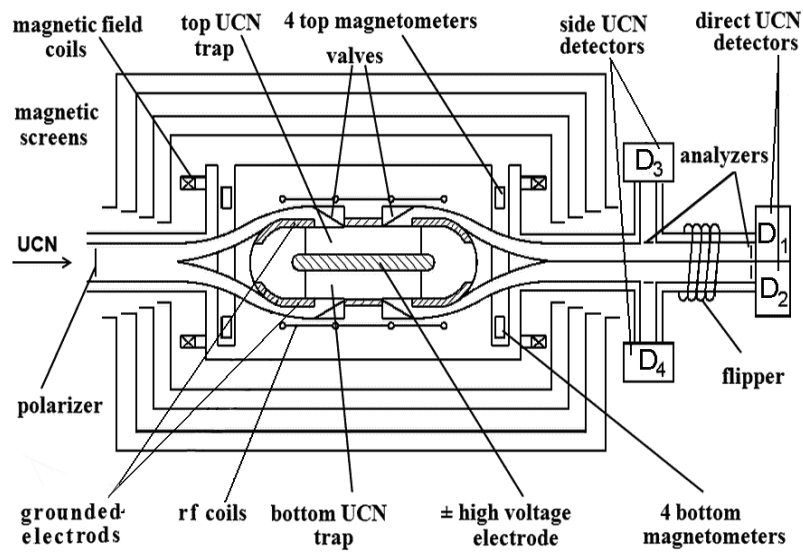

**Figure 1: Schematic side view (cut through the center) of the EDM spectrometer.**

A great advantage of using a symmetric double chamber is the possibility to compensate common drifts and fluctuations. On changing the polarity of the electric fields the shifts of precession frequencies due to the neutron EDM will have opposite signs in the two chambers, whereas those shifts due to common magnetic field instability will have the same signs and therefore do not affect the result for the neutron EDM. Another peculiar feature is a simultaneous analysis of both neutron polarization components with respect to the magnetic guide field. It is performed with two detectors attached to each storage chamber, each with its own polarization analyzer and spin flipper. Compared to a single-detector setup this increases total neutron counts by almost a factor two and in addition may be used for compensation of neutron intensity fluctuations from the source.

From the counts of each of the detectors $D_1 \ldots D_4$ one can derive corresponding experimental values $d_1 \ldots d_4$ for the neutron EDM. Compensation, respectively, determination of some systematic effects is obtained by using linear combinations of the individual values $d_i$:

$$EDM = \frac{1}{4}[(d_1 + d_2) + (d_3 + d_4)]$$

$$\Delta \nu = \frac{1}{4}[(d_1 - d_2) + (d_3 - d_4)]$$

$$\Delta N = \frac{1}{4}[(d_1 - d_2) - (d_3 - d_4)]$$

$$Z = \frac{1}{4}[(d_1 + d_2) - (d_3 + d_4)]$$

(3)

The first equation determines a value for the neutron EDM. For a fully symmetric setup, this combination of values $d_i$ fully compensates fluctuations common to both chambers and correlated with reversal of the electric fields. Moreover, the linearly independent second and third combinations in eq. (3) provide measurements of such fluctuations. While $\Delta \nu$ determines the effect of electric field influence on resonance conditions, $\Delta N$ measures a systematic effect on neutron count rates. Finally, in the last combination $Z$ all aforementioned effects (including the neutron EDM) are compensated, so that the condition $Z = 0$ provides a crucial test of the compensation scheme.

In order to perform measurements of the values $d_i$ one first has to determine the working point where the frequencies of free neutron precession and the rf field are exactly equal. This is done with neutron measurements for different time intervals $T$ between rf pulses, from which the working point is independent. Highest sensitivity to the neutron EDM is achieved if one applies phase shifts $\Delta\varphi$ of 90° or 270° between the two rf pulses. As shown in Fig. 2 this leads to largest slope $\partial N/\partial f$ of neutron counts at the working point (compare with eq. (2)). For each of the two phase shift settings his slope is determined from two neutron count rates obtained with small shifts of the radio-frequency by about $\Delta f = \pm 5 \times 10^{-4}$ Hz around the working point. The fact that the signs of the slopes are different for $\Delta\varphi = 90°$ and $\Delta\varphi = 270°$ is very useful to monitor drifts of the resonance curve which is used to provide additional stabilization of resonance conditions (note that for detecting influences of electric field polarity alteration on the neutron count rate (i.e. $\Delta N$ in eq. (3)) measured rates for $\Delta\varphi = 90°$ and $\Delta\varphi = 270°$ must be subtracted rather than added to account for the sign of the slope).

Measurements of the neutron EDM are performed within a sequence of different settings of electric field polarity, $\Delta\varphi$ and $\Delta f$. This sequence is chosen with the goal to minimize unwanted effects of drifts of the resonance condition. For instance, with $\Delta\varphi$ and $\Delta f$ in a given state the electric field polarity is alternated according to a sequence (+− −+) or (−++−) which eliminates the effect of linear drifts if measurements are performed within constant time intervals.

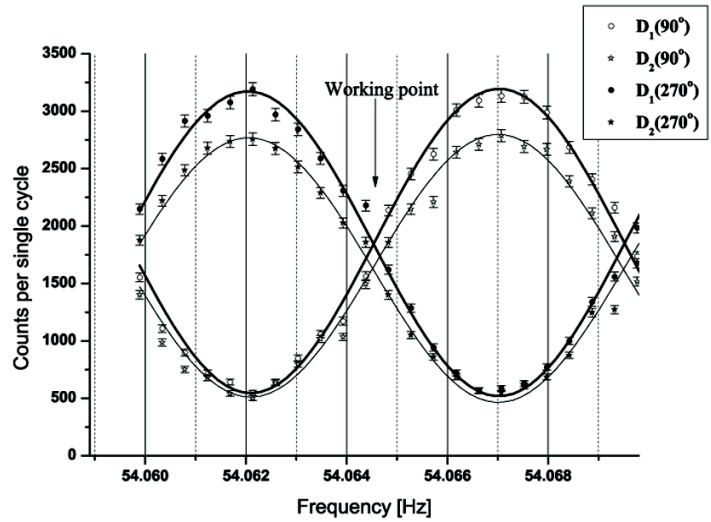

Figure 2: Central parts of Ramsey resonance curves for detectors $D_1$ and $D_2$ and phase shifts 90° and 270° between rf pulses. The visibility $\alpha$ of the resonance was 0.7 for $T = 95$ s.

All measurements were carried out at the ultracold neutron facility PF2 at the ILL in Grenoble, France. As determined in prior experiments the beam port PF2/MAM provides a unpolarized UCN density of about 7.5 n/cm$^3$ at the entrance of the EDM spectrometer [11]. Figure 3 presents an exemplary series of experimental data taken during 15 hours, with an electric field of 18 kV/cm and a UCNstorage time of 100 s. Each point represents the result of a single measurement sequence of the value of *EDM* in accordance with eqs. (1) and (3). Also quoted are separate results for the quantities $EDM_{top} = (d_1+d_3)/2$ and $EDM_{bott} = (d_2+d_4)/2$ for the top and bottom chambers. Total results from the series shown in Fig. 3 are $EDM_{top} = (2.59\pm3.90)\times10^{-25}$ e·cm, $EDM_{bott} = -(3.98\pm4.22)\times10^{-25}$ e·cm and $EDM = -(0.70\pm2.17)\times10^{-25}$ e·cm. The latter permits us to assess the sensitivity of the experiment when running under smooth conditions, which for this series of measurements amounted to $1.7\times10^{-25}$ e·cm/day.

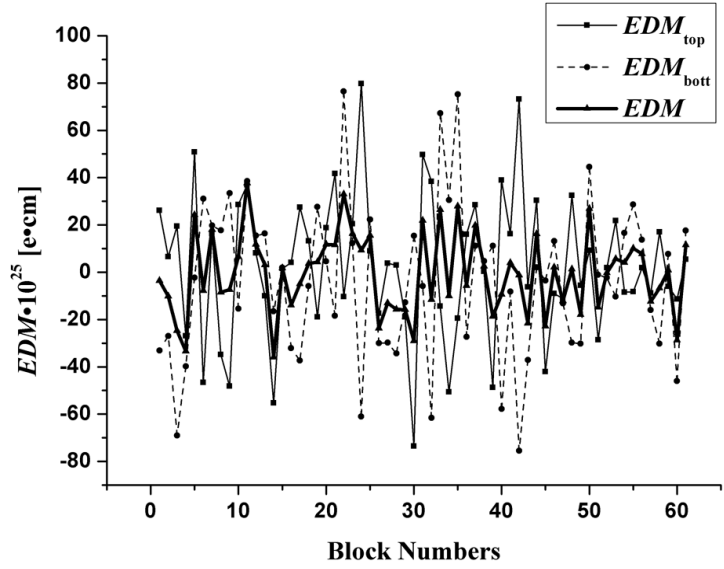

Figure 3: Exemplary series of measurements. $EDM_{top}$ and $EDM_{bott}$ are values measured for the top and bottom chamber separately (see text). $EDM$ is the quantity defined in eq. (3) that measures the neutron EDM using measurements with both chambers and all 4 UCN detectors.

Several of such series as shown in Fig. 3 were obtained at the ILL during three reactor cycles of 50 days each. Since after these measurements the reactor has been shut down for an extensive period of ten months for maintenance and upgrades, we felt it timely to present our first results. Results for $EDM$ are quoted in Table 1 together with measured values of the quantities $\Delta \nu$, $\Delta N$ and $Z$ defined in eq. (3), all in units of $10^{-26}$ e·cm. The deviations from zero of the values $EDM$ and $Z$ do not exceed one standard deviation. The values of $\Delta \nu$ and $\Delta N$ on the other hand deviate from zero by two and 3.5 standard deviations, the latter with different signs in the two measurements.

Table 1: Results of measurements in units of $10^{-26}$ e·cm.

|   | previous (PNPI) [4] | new (ILL) | total |
|---|---|---|---|
| $EDM$ | 0.7±4.0 | 0.36±4.68 | 0.56±3.04 |
| $\Delta \nu$ | -22.8±9.2 | -10.04±5.98 | -13.8±5.01 |
| $\Delta N$ | -14.5±4.4 | 18.62±5.15 | -0.53±3.35 |
| $Z$ | -0.8±4.0 | 3.68±4.72 | 1.05±3.05 |

Since the crucial criterion $Z = 0$ is fulfilled, we assume systematic effects visible in $\Delta \nu$ and $\Delta N$ as compensated in the value $EDM$. However, obviously our measuring scheme cannot control all possible systematic errors. In particular, local alterations of the magnetic field associated with leakage currents may escape detection by an assembly of a few discrete cesium magnetometers. In the reported measurements the leakage current during a typical data set was a few tens of nA and it did never exceed 2000 nA. As visible in Fig. 4 there is no obvious correlation between *EDM* and leakage current.

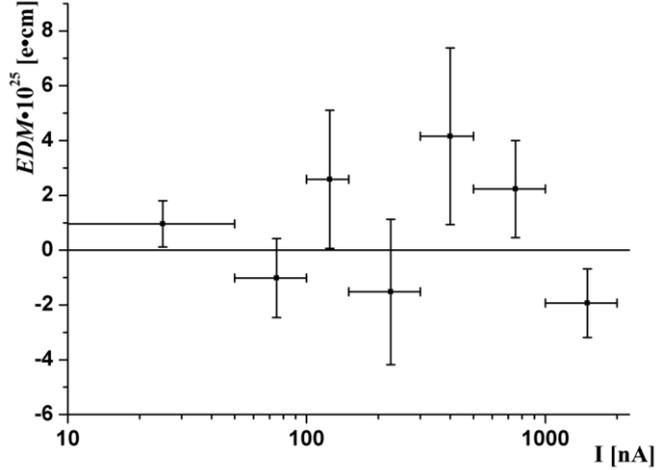

**Figure 4: Results of EDM measurements for different intervals of leakage current**

A separate control of possible effects on the cesium magnetometers has shown that after switching the electric field the difference in recordings of the magnetometers in the upper and lower planes (see Fig. 1) did not exceed 1 fT determined with all data accumulated during the EDM measurements. This limit can be converted into a limit of a corresponding false EDM effect of $2 \times 10^{-27}$ e·cm, i.e. by still more than a factor ten below the quoted accuracy of measurements.

Distributions of measured normalized values of *EDM*, $\Delta \nu$, $\Delta N$ and *Z* are presented in Fig. 5. (normalized values are $Y_i = \dfrac{y_i - <y>}{\sigma}$, where $y_i$ are measured values, $<y>$ is mean value, $\sigma$ is the standard deviation). The width of the distribution of values *EDM* and *Z* corresponds to that of a normal distribution defined by counting statistics. The distributions of values of $\Delta \nu$ and $\Delta N$ are somewhat broadened due to imperfect stability of magnetic field and neutron intensity. The absence of any broadening of the *EDM* and *Z* distributions demonstrates the compensation of fluctuations of the magnetic field.

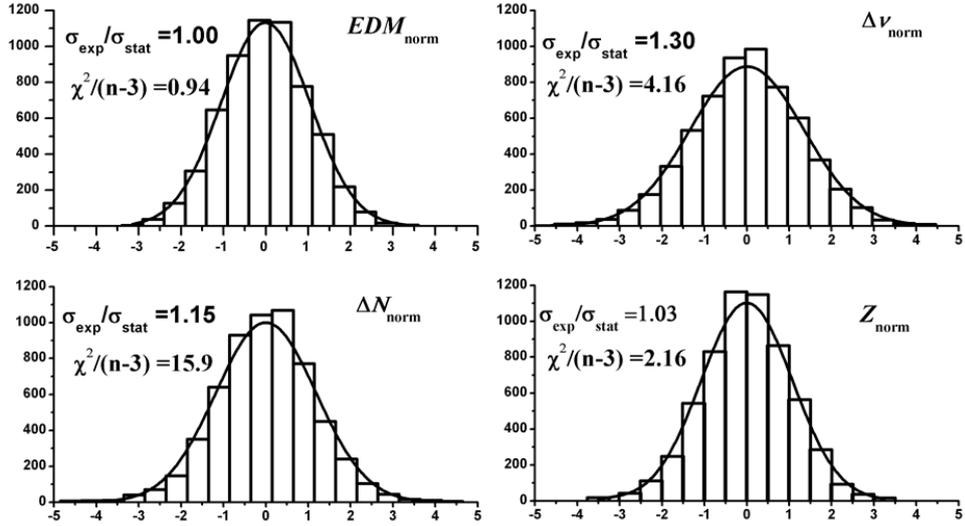

**Figure 5: Distribution of measured values for the quantities defined in eq. (3). For each quantity is shown the ratio of the width of the measured distribution to its uncertainty defined by counting statistics, as well as $\chi^2$ obtained from fitting a normal distribution.**

The accuracy of the reported measurements at ILL was $4.7 \times 10^{-26}$ e·cm. Including the earlier result obtained at the WWR-M reactor at PNPI [4] the total accuracy is $3.0 \times 10^{-26}$ e·cm. We interpret our new result as a limit on the neutron electric dipole moment of $|d_n| < 5.5 \times 10^{-26}$ e·cm at 90% confidence level. This experiment involves different systematic effects than other neutron EDM experiments. Notably it provides a result free from false effects due to geometric phases.

For upcoming runs of the experiment we plan to relocate the apparatus to another UCN beam which should bring a 3 to 4 fold gain in intensity at the entrance of the spectrometer. In addition, some redesign of the neutron transport within the spectrometer should provide a further increase of neutron count rates. With these measures in place the accuracy might be improved in near future down to $10^{-26}$ e·cm. Progress beyond that level will require novel UCN sources such as the upgrade of the source described in [12] projected at the ILL, or in a later stage a new source at the reactor PIK with prior implementation of a prototype at the WWR-M reactor at PNPI [13].

Concluding, the authors would like to thank the members of the subcommittee of college 3 for continuous support and their recommendation to perform neutron EDM measurements over an extended period of 2 – 3 years. They also express their gratitude to the staff of the workshop of experimental facilities in PNPI for their assistance in reconstructing the spectrometer, and to the personnel of the ILL reactor for help in assembling the installation and doing measurements. Particularly we would like to thank T. Brenner from ILL and A.I. Egorov, E.P. Volkov, A.N. Murashkin, E.V. Siber, A. Sumbatian and T. Savelieva from PNPI for their assistance. We also acknowledge a strong contribution of I. Tokmakov, N. Mikulinas and A. Pikalev on construction and operation of electronics and high-voltage source.